**Double well ground state energy splitting (or instanton flipping rate); rendering the implicit explicit.**

J.H. Hannay
H.H. Wills Physics Laboratory,
Tyndall Avenue,
Bristol BS8 1TL,
UK

J.H.Hannay@Bristol.ac.uk

Abstract
A prime example of quantum tunnelling is the semiclassical 'energy splitting' of the levels of symmetrical double well potential, or equivalently the flipping rate of an instanton. Curiously the accepted expression for the ground state splitting in terms of the (smooth) potential function has not been pursued to the full explicitness available from classical mechanics. This implicitness is rectified here.

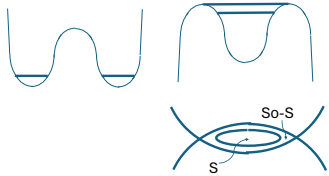

Fig 1. Left: a symmetric double well potential with classical motions of low energy $\hbar\omega/2$ indicated. The quantum ground state (a symmetric wavefunction) would have an energy slightly below this, and the next state (antisymmetric) equally above. Right: this quantum tunnelling 'splitting' is naturally expressed in terms of classical motion in an upturned potential, area $S$ in phase space. Also involved is the nearby peak-to-peak motion, the lemon shaped 'separatrix' of area $S_0$.

The symmetric double well potential provides a prime example of semiclassical quantum tunnelling. The first excited state energy is only slightly higher than the ground state energy – their energies are 'split' apart due to tunnelling (or weak coupling of the two individual wells). Curiously the accepted splitting formula in terms of the (smooth) potential function has not been pursued to the full explicitness available from classical mechanics. This implicitness[1] is rectified here. (Alternatively expressed in terms of instantons, the average flipping rate due to tunnelling is implicit since it equals the energy splitting, divided by $2\pi\hbar$).

An expression for the lowest two energies is long known (an outline derivation is given in the final 'appendix' paragraph below):

$$\tfrac{1}{2}\hbar\omega \pm \tfrac{1}{2}\sqrt{\tfrac{\pi}{e}}\tfrac{\hbar\omega}{\pi}\exp\left(-\tfrac{S}{2\hbar}\right) \qquad (1)$$

Here $\omega$ is explicit: it is a classical oscillation angular frequency $\omega = \sqrt{mV''}$ where $V''$ is the second derivative of potential $V(q)$ at the well bases (where $V=0$). However what has been lacking is a proper formula for $S$ in the 'tunnelling exponential' $\exp(-S/2\hbar)$. $S$ is famously expressed geometrically as the phase space area, the action, of the classical to-and-fro motion in an *upturned* potential (fig 1) at an energy $E = \tfrac{1}{2}\hbar\omega$ *below* the twin peaks (the upturned well bases). It has hitherto been left as an implicit expression, but an explicit formula is available for $S$ as follows. (The formula for $S$, (2) with (4), actually applies unchanged for semiclassical excited states too, but the prefactors of the exponential in (1) are different[2,3], and the quantized energies $E$ themselves are not fully explicit, in contrast to the ground state).

$S$ is slightly less than $S_0$, the area of the separatrix (which is explicitly given by $S_0 = 2\int\sqrt{2mV(q)}dq$ with integration limits at the double well bases). In fact

$$S \approx S_0 + \tfrac{2E}{\omega}\log\left(\tfrac{E}{e\varepsilon}\right) \qquad (2)$$

leading to

$$T = -\tfrac{dS}{dE} \approx -\tfrac{2}{\omega}\log\left(\tfrac{E}{\varepsilon}\right) \qquad (3)$$

for the time period of the orbit needed shortly. Here $\varepsilon$ is an as-yet-unknown constant with the dimensions of energy; it is to be found here, rendering the energy splitting explicit. The result is this, where $P$ is the central momentum $\sqrt{2mV_{max}}$.

$$\varepsilon = \left(\tfrac{2P^2}{m}\right)\exp\left[2\omega\int_0^P \tfrac{m}{p}\left(\left(\tfrac{dq}{dp}\right)_{sep} - \tfrac{1}{m\omega}\right)dp\right] \qquad (4)$$

Substituting (4) into (2), and (2) into (1) yields the explicit energy splitting. Finding the time period $T$, (3), by an alternative argument now, will derive (4).

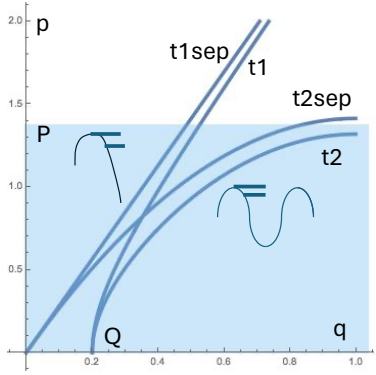

Fig 2  Phase space (q,p), showing a quarter period motion, with duration t2 that is sought, and three associated motions, also while p<P (the shaded box), that can be used to find t2.

Consider one quarter of the upturned *double* well orbit starting from p=0, calling its start position Q say, and ending at its maximum momentum P (fig 2). The sought duration *t2* for this (a quarter of the period) is to be compared with the known duration *t1* for that same interval of p and starting from the same position Q of the upturned *single* well harmonic oscillator. Corresponding 'separatrix' durations *t2*$_{sep}$ and *t1*$_{sep}$, with release from an upturned peak point (double or single) and covering the same p interval 0 to P, are both infinite. But crucially their difference *t2*$_{sep}$ − *t1*$_{sep}$, is finite and calculable below, (6). The key fact is then that in the semiclassical limit, *t2 − t1 = t2*$_{sep}$ − *t1*$_{sep}$.

The 0 to P duration *t1* is available from the action $\int_0^P q\, dp$ with the hyperbola equation $m\omega^2 q^2/2 = p^2/2m + m\omega^2 Q^2/2$. This action then needs to be differentiated with respect to the energy $m\omega^2 Q^2/2$ to obtain the exact duration of this segment of hyperbola. It involves several terms with logs and square roots, but its leading term in the relevant limit Q→0 is simply

$$t1 = \frac{1}{\omega}\log\left(\frac{2P}{m\omega Q}\right) = \frac{1}{\omega}\log\left(\frac{2P}{\sqrt{m\omega\hbar}}\right) \tag{5}$$

where the appropriate value of Q has been given by energy matching: $m\omega^2 Q^2/2 = \hbar\omega/2$. The other ingredient, *t2–t1*, equals, thanks to the 'key fact' above,

$$t2_{sep} - t1_{sep} = \int_0^P \frac{m}{p}\left(\left(\frac{dq}{dp}\right)_{sep} - \frac{1}{m\omega}\right)dp \tag{6}$$

The reasoning for the first integrand term here is that the phase space velocity along the separatrix has a component p/m along the horizontal q direction, and therefore a component (p/m)(dp/dq) along the vertical p direction. The reciprocal of this latter has, subtracted from it, the corresponding reciprocal for the upturned harmonic oscillator (straight separatrix). Their difference has finite integral (its only singularity is an inverse square root one at p=P).

The semiclassical time period for the whole orbit, rather than a quarter, is then 4×((5)+(6)), and this needs to be equated to (3), with $E = \frac{1}{2}\hbar\omega$, to supply the desired energy constant $\varepsilon$ (4), for substitution into (2) and finally into (1).

$$T = -\frac{2}{\omega}\log\frac{\hbar\omega}{2\varepsilon} = \frac{4}{\omega}\log\left(\frac{2P}{\sqrt{m\omega\hbar}}\right) + 4\int_0^P \frac{m}{p}\left(\left(\frac{dq}{dp}\right)_{sep} - \frac{1}{m\omega}\right)dp \tag{7}$$

(For the simple quartic potential $(q^2-L^2)^2$ the integral in (7) is straightforward, and reproduces the result of a sophisticated asymptotic path integral evaluation[5] of the energy splitting that is possible in that case.)

This final 'appendix' paragraph is an outline derivation of (1). Mathematically $\mathcal{N}\, k(q)^{-1/2}\exp\left[-\int_{\sqrt{\hbar/m\omega}}^q k(q)\,dq\right] \to (m\omega/\pi\hbar)^{1/4}\exp\left[-m\omega q^2/2\hbar\right]$ as $q\to\infty$ where $k(q) = \sqrt{2m(m\omega^2 q^2/2 - \hbar\omega/2)}/\hbar$ and $\mathcal{N}^2 = m\omega/\hbar\sqrt{4\pi e}$. The right side is the exact ground state wavefunction of a harmonic oscillator. Therefore the left side, with the $\mathcal{N}$ as given, is the correct semiclassical (WKB) wavefunction tail for this ground state. Consider the same left side, with the *same* $\mathcal{N}$, but with the double well potential (with a minimum at q=0, and $\omega = \sqrt{mV''}$ there) replacing the $m\omega^2 q^2/2$ in $k(q)$. This new left side has a value at the middle (peak potential), written schematically $\mathcal{N}\, k^{-1/2}\exp(-\int k)$, and a semiclassical derivative $-\mathcal{N}\, k^{+1/2}\exp(-\int k)$ there. By Herring's formula[3,4], the energy splitting is proportional to their product, specifically $\Delta E = 2(\hbar^2/m)\mathcal{N}^2\exp\left(-2\int k\right)$, which is the splitting in (1). (Factors of 4 from double contributions, and 1/2 from normalization, have been accounted for).

*Acknowledgement*

My thanks to Michael Berry for a confirmatory calculation for the special case of a simple quartic well (S is then given exactly by 'complete elliptic integral' functions).